\documentclass[pra,aps,reprint,showpacs,amsmath,superscriptaddress]{revtex4-1}

\usepackage{bm}
\usepackage{graphicx}
\usepackage{dsfont}

\usepackage{bm}
\usepackage{mathrsfs}

\usepackage{framed}

\usepackage[all]{xy}
\usepackage{graphicx}
\usepackage{framed}
\usepackage{comment}
\usepackage{here}                    
\usepackage{latexsym}		     
\usepackage{amsmath}     
\usepackage{amsfonts}  
\usepackage{amssymb}
\usepackage{amsthm}
\usepackage{dcolumn}
\usepackage{color}
\usepackage{mathrsfs}

\newtheorem{example}{Example}

\newtheorem{definition}{Definition}

\renewcommand{\phi}{\varphi}
\renewcommand{\>}{\rangle}
\newcommand{\<}{\langle}

\newcommand{\ket}[1]{|#1\>}
\newcommand{\bra}[1]{\<#1|}

\newcommand{\be}{\begin{equation}}
\newcommand{\ee}{\end{equation}}
\newcommand{\bea}{\begin{eqnarray}}
\newcommand{\eea}{\end{eqnarray}}

\newcommand{\Int}{\mathbb{Z}}

\renewcommand{\phi}{\varphi}

\usepackage[colorlinks,citecolor=blue,linkcolor=blue,urlcolor=blue]{hyperref}

\begin{document}

\title{Decision and function  problems based on boson sampling}

\author{Georgios M. Nikolopoulos}
\email{nikolg@iesl.forth.gr}

\affiliation{Institute of Electronic Structure \& Laser, 
FORTH, P.O. Box 1385, GR-70013 Heraklion, Greece}

\author{Thomas Brougham}
\affiliation{School of Physics and Astronomy, University of Glasgow, Glasgow G12~8QQ, United Kingdom}

\date{\today}

\begin{abstract}
Boson sampling is a mathematical problem that is strongly believed to be intractable for classical computers, whereas passive linear interferometers can produce samples efficiently.
So far, the problem remains a computational curiosity, and the possible usefulness of boson-sampling devices is mainly limited to the proof of quantum supremacy.  The purpose of this work is to investigate whether boson sampling can be  used as a resource of decision and function 
problems that are computationally hard, and may thus have cryptographic applications. After the definition of  a rather general theoretical framework for the design of such problems, we discuss their solution by means of  a 
brute-force numerical approach, as well as by means of non-boson samplers.  
Moreover, we estimate the sample sizes required for their solution by passive linear interferometers, and it is shown that they are independent of the size of the Hilbert space. 
\end{abstract}


\maketitle


\section{Introduction} 
\label{sec1}

In the quest for solid examples of quantum supremacy, boson sampling 
(BS) has a prominent place. The first reason for this, is that the classical simulation of BS  pertains to the computation of permanents of large matrices with complex entries; a problem known to be computationally hard  \cite{complex1,complex2,complex3}.   
In fact, the existence of an efficient classical algorithm for simulating exact (and presumably  even approximate) BS would have severe implications, which are strongly believed to be highly implausible  in  computational complexity theory \cite{complex1,complex2,complex3}. 
The second reason is that, contrary to classical computers, networks consisting of passive linear optical elements (beamsplitters and phase-shifters) only \cite{complex1,BSintro}, can efficiently produce samples of unknown boson distributions.
The set-up also requires single-photon sources, and photodetectors, but no  quantum memories or feedback \cite{complex1,BSintro}. 

The first experiments on small-scale BS problems have confirmed many of the theoretical predictions \cite{ExpBS1,ExpBS2,ExpBS3,ExpBS4,ExpBS5,ExpBS6,ExpBS7}.  
A proof of quantum supremacy, however, requires considerably larger-scale BS (typically around $N\sim 30$ identical photons in a network with $\sim N^2$ modes) \cite{complex1}. Implementation of 
large-scale BS may require significant effort,  but by contrast to universal quantum computers that remain a formidable challenge, it is within reach of today's technology. In any case, a solid proof of quantum supremacy, has to be accompanied by additional scientific evidence for the verification of the operation of the BS device. Unfortunately, classical simulation of BS 
 becomes very quickly impossible as one increases the complexity of the problem, and thus the verification by classical means 
 is not possible.  
Moreover, although BS devices are capable of producing samples efficiently, the reconstruction of the entire probability 
distribution would require an exponentially large number of measurements, which is not possible in practice. 
Hence, over the last years considerable literature has been devoted to the development of witnesses, which can be used for the  efficient verification of BS devices, and can rule out alternative models or mock-up distributions \cite{ExpBS6,ExpBS7,NonUni,TichyPRL14,wang16,genBunching}. 

Modern cryptography relies on computationally hard {\em function problems}, such as  integer factorization and discrete logarithm, which 
are intimately connected to {\em decision problems} (e.g., see chapter 9 of \cite{handbook}).
The probabilistic nature of BS 
complicates considerably its connection to decision/function problems, 
and thus, for the time being, the usefulness of BS devices is limited to proof-of-principle demonstrations of quantum speed-up 
\cite{complex1,complex2,complex3,BSintro,ExpBS1,ExpBS2,ExpBS3,ExpBS4,ExpBS5},  
or  to quantum simulations \cite{BS-sim}, and there are no known real applications. More precisely, it is not clear whether such a connection is possible, 
or what type of decision/function problems may be of relevance \cite{complex1,BSintro}. 

In the present work we address these open questions,  by defining a rather 
general theoretical framework for decision/function problems that rely on BS. 
In an attempt to establish a clear connection between BS and  the   proposed class of problems, 
we investigate in detail how the key quantities of the problems depend 
on the BS parameters, i.e., on  the photon configuration at the input and the unitary of the network.  Subsequently, the  solution of the problems is discussed in the framework of a brute-force numerical approach that employees Ryser's algorithm. Our estimates suggest that such an approach becomes very quickly  intractable due to the exponentially large number of permanents 
involved in the calculation. A solution by means of non-boson samplers is also not in general possible, whereas the problems under consideration can be solved efficiently (in terms of the required sample sizes),  if one is capable of performing BS experiments.  

The paper is organized as follows. 
Section  \ref{sec2} is devoted to preliminaries, whereas in Sec. \ref{sec3} we discuss the binning of boson-sampling data. Binning plays a pivotal role in our theoretical framework  in Sec. \ref{sec4}, which allows for the definition 
of a class of decision and function problems. A summary with concluding remarks is given in Sec. \ref{sec5}.


\section{Preliminaries}
\label{sec2}

BS is  the generalization of the standard Hong-Ou-Mandel effect   
to a network with $M$ modes and $N$ photons \cite{complex1,complex2,complex3,BSintro}. 
The network is 
governed by an $M\times M$ unitary $\hat{\mathfrak U}$, with complex matrix elements $U_{m,n}$, which is chosen randomly according to the Haar measure. In the absence of losses, the evolution of the system 
is restricted to the Hilbert space ${\mathbb H}_{M,N}$ 
spanned by the orthonormal basis states $\{\ket{{\bm t}_i}\}$ with the $i$th 
state defined as 
\[
\ket{{\bm t}_i} :=  (\hat{a}_1^{\dag} )^{t_{i,1}}\ldots (\hat{a}_M^{\dag} )^{t_{i,M}} \ket{{\bm 0}},\,  t_{i,j}\in {\mathbb Z}_{N+1}\] and ${\mathbb Z}_{q}:=\{0,1,\ldots,q-1\}$. The creation operation  $\hat{a}_j^{\dag}$ creates a photon at the $j$th mode, and for the $i$th sate we have 
\bea
\sum_{j=1}^M t_{i,j} = N.
\label{cons:eq}
\eea
The tuple 
\bea
{\bm t}_i := (t_{i,1},t_{i,2},\ldots, t_{i,M}),
\label{tuple}
\eea
refers to the photon configuration at the  $i$th state $\ket{{\bm t}_i}$, and let 
\bea
{\mathbb S}_{M,N}:=\{{\bm t}_1, {\bm t}_2,  \ldots  {\bm t}_{|{\mathbb S}_{M,N}|}\}
\label{set:eq}
\eea
denote 
the  set of all possible configurations.  The total number of different photon  configurations $|{\mathbb S}_{M,N}|$,  is given by 
the binomial coefficient 
\bea
 \binom{M+N-1}{N}.
 \label{H-size}
 \eea

\begin{example}
\label{example1}
Consider the case of $N=2$ photons in a network of $M=4$ modes. The Hilbert space is spanned by $10$ basis states: 
$\ket{{\bm t}_1}=\ket{2,0,0,0}$, 
$\ket{{\bm t}_2}=\ket{1,1,0,0}$, 
$\ket{{\bm t}_3}=\ket{0,2,0,0}$, 
$\ket{{\bm t}_4}=\ket{1,0,1,0}$, 
$\ket{{\bm t}_5}=\ket{0,1,1,0}$, 
$\ket{{\bm t}_6}=\ket{0,0,2,0}$, 
$\ket{{\bm t}_7}=\ket{1,0,0,1}$, 
$\ket{{\bm t}_8}=\ket{0,1,0,1}$,  
$\ket{{\bm t}_9}=\ket{0,0,1,1}$, 
and 
$\ket{{\bm t}_{10}}=\ket{0,0,0,2}$.  In state $\ket{{\bm t}_1}$, 
both photons occupy the first mode, whereas for state $\ket{{\bm t}_7}$, 
the photons occupy the first and the fourth mode. Hence, the set of all possible configurations  is ${\mathbb S}_{4,2}=\{{\bm t}_1=(2,0,0,0); {\bm t}_2=(1,1,0,0); \ldots;{\bm t}_{10}= (0,0,0,2)\}$.
\end{example} 
 
The configuration (\ref{tuple}) is basically the $(N+1)$-adic expansion of the integer 
\bea
\tau_i := \sum_{j=0}^{M-1} t_{i,j+1}N^j,  
\label{weight:eq}
\eea 
with the digits (elements) $t_{i,j}$ satisfying Eq. (\ref{cons:eq}). 
For a given pair $\{M,N\}$, there is a one-to-one correspondence between the set of possible configurations (\ref{set:eq}), and the set of their corresponding integer representations $\{\tau_1, \tau_2,\ldots,\tau_{|{\mathbb S}_{M,N}|}\}$.  
Hence, the integer representation facilitates the unambiguous ordering of all possible configurations in the set ${\mathbb S}_{M,N}$, according to a prescribed rule e.g., in ascending order.  

\begin{example} 
The integer representations of the tuples (configurations) pertaining to the states of example \ref{example1} are the following: 
$\tau_1 = 2$, $\tau_2 = 3$, $\tau_3 = 4$, 
$\tau_4 = 5$, $\tau_5 = 6$, $\tau_6 = 8$,
$\tau_7 = 9$, $\tau_8 = 10$, $\tau_9 = 12$,
$\tau_{10} = 16$. Hence, the states (configurations) of example  \ref{example1} are arranged in ascending order according to their 
integer representations.  
\end{example}

Consider now a photon configuration ${\bm s}$, chosen at random from a uniform distribution over the set ${\mathbb S}_{M,N}$. 
The  basis state $\ket{\bm s}$ is prepared at the input of the $M-$mode network, and its evolution is determined by the unitary  map 
\[
\hat{\mathfrak U} \hat{a}_j^\dag \hat{\mathfrak U} ^\dag \to \sum_{j^\prime=1}^{M} U_{j,j^\prime} \hat{a}_{j^\prime}^\dag. 
\]
The output state is a superposition of the form 
\bea
\ket{\Psi_{\rm out}^{({\bm s};\hat{\mathfrak U})}}=\sum_{{\bm r} \in{\mathbb S}_{M,N}} C_{{\bm s},{\bm r}} \ket{\bm r}; 
\quad C_{{\bm s},{\bm r}}:=\bra{\bm r}\hat{\mathfrak{V}}\ket{\bm s},
\label{Psi_out:eq}
\eea
where $\hat{\mathfrak{V}}$ is a unitary acting on the $N$-photon space and is related to $\hat{\mathfrak{U}}$  by a natural homomorphism \cite{complex1}. The output photon configuration  ${\bm r}$ is a 
random string,  which takes values from the set ${\mathbb S}_{M,N}$. That is 
${\bm r}$ can take one of the $|{\mathbb S}_{M,N}|$ different possible forms 
$\{{\bm t}_1, {\bm t}_2,\ldots,{\bm t}_{|{\mathbb S}_{M,N}|}\}$, with probabilities 
\bea
P({\bm t}_{i}|{\bm s};\hat{\mathfrak U}) = |C_{{\bm s},{\bm t}_i}|^2,
\label{Pc:eq}
\eea
 and 
\bea
\sum_{{\bm r}\in{\mathbb S}_{M,N}} P({\bm r}|{\bm s};\hat{\mathfrak U})  = 1.
\label{BS:norm}
\eea 
It can be shown that the complex  amplitudes $C_{{\bm s},{\bm r}}$ are directly related to the permanents of 
$N\times N$ sub-matrices of $\hat{\mathfrak U}$, denoted by  $\hat{\mathfrak U}_{{\bm s},{\bm r}}$ \cite{complex1,complex2,BSintro,Scheel,valiant} i.e.,  $C_{{\bm s},{\bm r}}\propto \textrm{Per}[\hat{\mathfrak U}_{{\bm s},{\bm r}}]$. 
The probability distribution of ${\bm r}$, given the input configuration ${\bm s}$ and fixed unitary $\hat{\mathfrak U}$ will be denoted by 
\bea
\bar{\cal P}_{\bm r}^{({\bm s};\hat{\mathfrak U})} := \{P({\bm r}|{\bm s};\hat{\mathfrak U})\,:\,{\bm r}\in {\mathbb S}_{M,N} \}.
\label{P_rs:eq} 
\eea 
  
We see therefore that BS is fully characterized by the unitary of the network $\hat{\mathfrak U}$  and input photon configuration ${\bm s}$ (to be referred to hereafter as {\em seed}). For sufficiently large values of $M,N\gg 1$, the simulation of a 
{\em BS session}  ${\tt BS}({\bm s},\hat{\mathfrak U})$, and the estimation of the corresponding distribution $\bar{\cal P}_{{\bm r}}^{({\bm s};\hat{\mathfrak U})}$, become an intractable problem for classical computers, because it requires the estimation of permanents for an exponentially large number of complex $N\times N$ matrices \cite{complex1,complex2,complex3,BSintro,valiant}.  

By contrast, one can sample efficiently from the unknown boson distribution 
$\bar{\cal P}_{{\bm r}}^{({\bm s};\hat{\mathfrak U})}$, in a  passive linear-optic network that implements 
$\hat{\mathfrak U}$ and using single-photon sources, and photodetectors \cite{BSintro,ExpBS1,ExpBS2,ExpBS3,ExpBS4,ExpBS5,ExpBS6,Scheel}.  
The sampling from $\bar{\cal P}_{{\bm r}}^{({\bm s};\hat{\mathfrak U})}$ pertains to a number  of experimental ``runs" (say ${\mathscr N}$). In each run, the same basis state $\ket{{\bm s}}$ is input in the network, where it evolves 
according to the implemented unitary $\hat{\mathfrak U}$. 
The quantum state at the output of the 
network is projected onto the $|{\mathbb S}_{M,N}|$ possible basis states, 
by means of 
coincidence photodetection at the output modes. In particular, the projective measurement returns  the $i$th photon configuration in the set (\ref{set:eq})  with probability $P({\bm t}_i|{\bm s};\hat{\mathfrak U})$. 
Hence, by performing the same experiment many times (i.e., for sufficiently large values of ${\mathscr N}$), one essentially can approximate all the probabilities $\{P({\bm t}_i|{\bm s};\hat{\mathfrak U})\}$ by the frequencies of occurrence (proportions) of the corresponding configurations $\{{\bm t}_i\}$, obtaining  an estimate for the distribution $\bar{\cal P}_{{\bm r}}^{({\bm s};\hat{\mathfrak U})}$. 
The faithful reconstruction of the entire BS distribution $\bar{\cal P}_{{\bm r}}^{({\bm s};\hat{\mathfrak U})}$ 
for given $\{{\bm s},\hat{\mathfrak U}\}$ requires exponentially large sample, and becomes impractical for large-scale BS i.e., 
for sufficiently large values of $M,N\gg 1$. 
Hence, a number of verification schemes have been proposed in the literature, 
which can verify the operation of a BS device and rule out alternative explanations \cite{ExpBS6,ExpBS7,NonUni,TichyPRL14}.  
 
The {\em dilute limit} of BS corresponds to $M={\cal O}(N^2)$ modes, where the probability of observing configurations with multiple occupancy of modes at the output is small \cite{complex1,BSintro,remark1}. 
From a theoretical point of view, in this case the problem of BS can be analysed accurately  in the framework of the $\binom{M}{N}$ different ``collision-free" configurations, and the effective  size of the Hilbert space scales as   \cite{complex1,complex2,complex3,BSintro} 
\bea
|{\mathbb H}_{M,N}|=|{\mathbb S}_{M,N}|\sim (MN^{-1})^N.  
\label{H-size2}
\eea 
Strictly speaking,  the 
computational hardness of BS has been proved rigorously only for networks where the number of modes scales quadratically with the number of photons.     
The computational hardness of BS, however, is believed to hold also for $M={\cal O}(N)$  (e.g., see Sec. 6.2 in \cite{complex1}). 

The  main ideas and results presented in the following section are expected to be valid for any  combination of $\{M,N\}$, but for the sake of compatibility with existing work in the field, we will mainly present results for cases 
where the number of modes scales quadratically with the number of photons.

\begin{figure}
\includegraphics[scale=0.82]{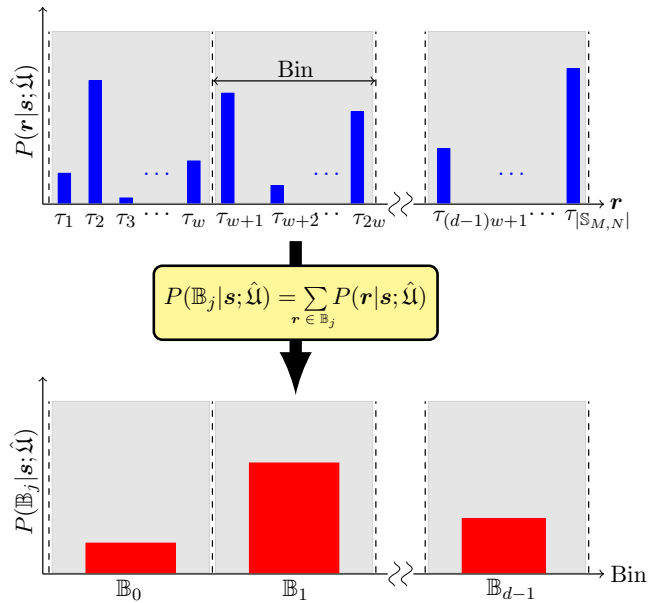}
\caption{(Color online) A schematic representation of binning. 
The outcomes of the BS distribution are assumed to be arranged in ascending order according to their integer representations (see Sec. \ref{sec2}).
}
\label{fig1}
\end{figure}


\begin{figure*}
\includegraphics[scale=0.3]{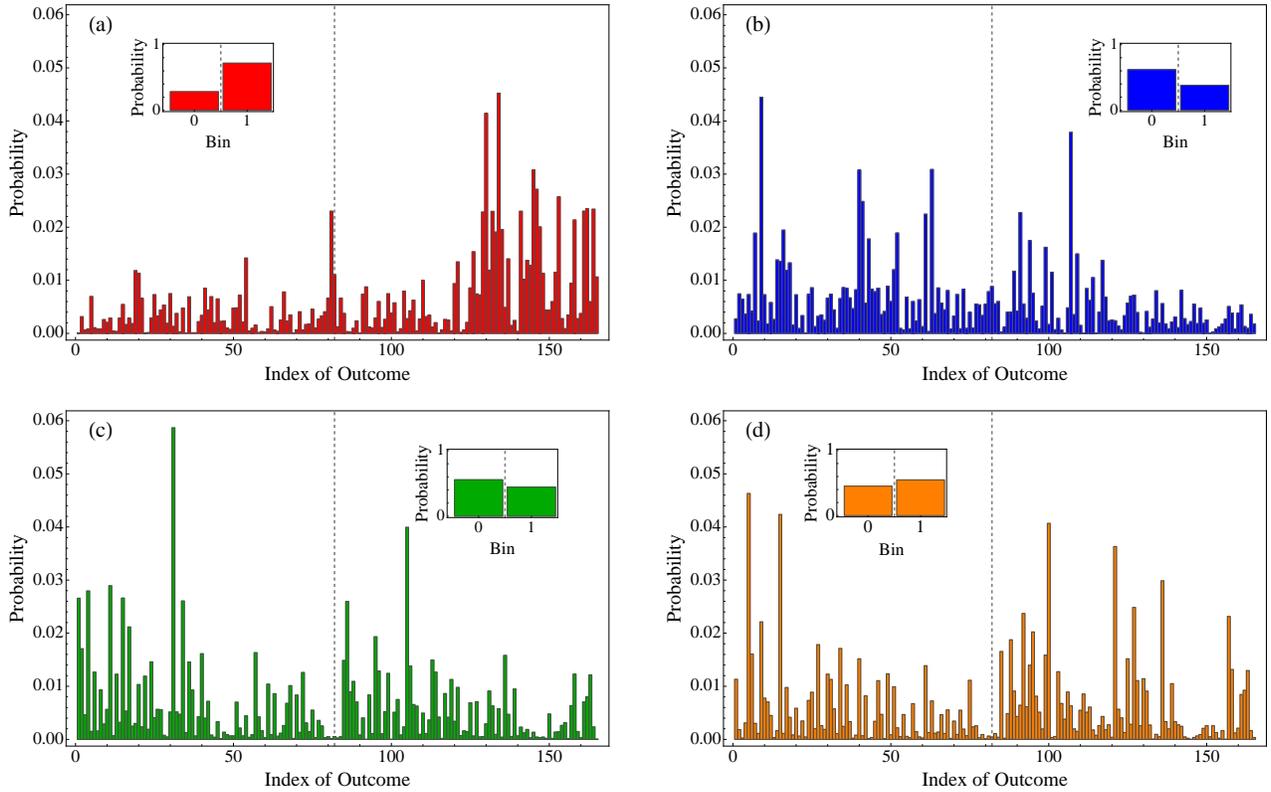}
\caption{(Color online) Illustration of binning with $d=2$ bins, for boson-sampling distributions $\bar{\cal P}_{\bm r}^{({\bm s};\hat{\mathfrak U})}$ pertaining to two different randomly chosen unitaries, and two different seeds. Plots on the same row correspond to the same unitary but different seeds, whereas plots on the same column refer to the same seed but different unitaries. 
The vertical dashed lines separate the outcomes that contribute to the two bins, and 
the insets show the corresponding distributions ${\cal P}_{\beta}^{({\bm s};\hat{\mathfrak U})}$ after the binning.  
Other parameters: 
$M=11$ and $N=3$.
}
\label{fig2}
\end{figure*}

\section{Binning of Boson-sampling data} 
\label{sec3}

Consider a BS session determined by the parameters $\{{\bm s},\hat{\mathfrak U}\}$. 
The set of all possible configurations (\ref{set:eq}) can be formally  divided into 
$d>1$ {\em disjoint} subsets $\{ {\mathbb B}_j~:~j\in \Int_d\}$, to be referred to 
hereafter as {\em bins}.  This {\em binning}  is purely classical and does not require any prior knowledge on the underlying BS 
distribution  $\bar{\cal P}_{\bm r}^{({\bm s};\hat{\mathfrak U})}$.  As long as one knows the number of photons and the number of modes, then ${\mathbb S}_{M,N}$ is known, and can be easily partitioned into $d$ subsets.  For reasons that will become evident below, we are interested in $d\ll |{\mathbb S}_{M,N}|$ bins.

There is no unique way of binning, but for the sake of simplicity throughout this work we adopt a rather simple approach (see Fig. \ref{fig1}). The configurations in the set ${\mathbb S}_{M,N}$ are assumed to be arranged in ascending order according to their integer representations [see Eq. (\ref{weight:eq})],  
and the $j$th bin pertains to the subset of configurations   
\bea
{\mathbb B}_j=\{{\bm t}_k\in{\mathbb S}_{M,N}~:~ T_{j}< k\leq T_{j+1}\};\quad  j\in\Int_d,
\label{bin:eq}
\eea 
with 
\begin{equation*}
T_j = \left\{
\begin{array}{cl}
0, & \text{if }\, j = 0\\
\sum_{i=0}^{j-1} w_i, & \textrm{otherwise}.
\end{array} \right.
\end{equation*}
The width of the $j$th bin is given by 
\bea
w_j:=\left \lfloor \frac{|{\mathbb S}_{M,N}|}{d} \right \rfloor+\Theta_j,
\label{w:eq}
\eea
where 
\bea
\Theta_j = \left\{
\begin{array}{cl}
1, & \text{if }\, j < q\\
0, & \textrm{otherwise},
\end{array} \right.
\label{w2:eq}
\eea
and 
$q<d$ is the remainder of the division 
$|{\mathbb S}_{M,N}| d^{-1}$.

For this particular definition,  all the bins have the same size when $q=0$, whereas 
for $q\neq 0$ they differ by at most one outcome; a difference which is not expected to be so prominent when $d\ll |{\mathbb S}_{M,N}|$.  
Each bin is uniquely identified by its label, 
and  for the sake of simplicity we can introduce a random discrete variable $\beta$ which takes values from the set of all possible $d$ labels i.e., $\beta\in \Int_d$. 

Having organized the possible outcomes into a number of disjoint bins, one may work with 
a new probability distribution  
\[
{\cal P}_{\beta}^{({\bm s};\hat{\mathfrak U})}:= \{ P({\mathbb B}_j|{\bm s};\hat{\mathfrak U})~:~ j\in \Int_d \},
\]
where the conditional probability of occupancy of the $j$th bin $P({\mathbb B}_j|{\bm s};\hat{\mathfrak U})$, 
is directly related to the corresponding conditional probabilities for the outcomes that 
contribute to the bin. 
To see this, consider the projector onto the subspace pertaining to the basis states with 
configurations from  $j$th bin i.e., 
\bea
\hat{\mathfrak{P}}_j := \sum_{{\bm b}\in{\mathbb B}_j}\ket{{\bm b}}\bra{{\bm b}}.
\eea
Using Eqs. (\ref{Psi_out:eq}) and (\ref{Pc:eq}), the probability for obtaining an outcome in the $j$th bin for given 
$\{{\bm s},\hat{\mathfrak U}\}$ is 
\bea
P({\mathbb B}_j|{\bm s};\hat{\mathfrak U}) =\bra{\Psi_{\rm out}^{({\bm s};\hat{\mathfrak U})}}\hat{\mathfrak{P}}_j\ket{\Psi_{\rm out}^{({\bm s};\hat{\mathfrak U})}}=\sum_{{\bm r} \in {\mathbb B}_j} 
P({\bm r}|{\bm s};\hat{\mathfrak U}), 
\label{probs:eq}
\eea 
with the normalization
\bea
\sum_{j=0}^{d-1} P({\mathbb B}_j|{\bm s};\hat{\mathfrak U})  = 1.
\label{norm:eq}
\eea
For the sake of illustration, in Fig. \ref{fig2} we show the binning for four different BS distributions $\bar{\cal P}_{\bm r}^{({\bm s};\hat{\mathfrak U})}$, 
corresponding to  two different unitaries and two different seeds, for $d=2$ bins. 


\subsection{The most-probable-bin function}

At this point we have all the necessary ingredients to define mathematical problems and functions pertaining to ${\cal P}_{\beta}^{({\bm s};\hat{\mathfrak U})}$.  

\begin{definition} 
\label{def1} For a given BS set-up,  the most-probable-bin function ${\tt MPB}_{d;\hat{\mathfrak U}}({\bm s})$ is defined as follows. \\
Parameters: Number of bins $d$, and the unitary $\hat{\mathfrak{U}}$ of the $M-$mode network. \\
Input: $N-$photon seed of BS ${\bm s}\in {\mathbb S}_{M,N}$. \\
Output: The most probable bin in ${\cal P}_{\beta}^{({\bm s};\hat{\mathfrak U})}$.
\end{definition}

For given $\{d,\hat{\mathfrak U}\}$, the function ${\tt MPB}_{d;\hat{\mathfrak U}}$ is a map 
\bea
{\tt MPB}_{d;\hat{\mathfrak U}}~:~{\mathbb S}_{M,N}\mapsto \Int_d.
\label{map:eq}
\eea
The output i.e., the image of the seed ${\bm s}\in {\mathbb S}_{M,N}$ under  ${\tt MPB}_{d;\hat{\mathfrak U}}$, is an element of $\Int_d$ i.e., a dit, and refers to the label of the most-probable bin.
Our task now is to discuss the dependence of  the most-probable bin of the distribution 
${\cal P}_{\beta}^{({\bm s};\hat{\mathfrak U})}$ on the number of bins, as well as on the unitary of the network $\hat{\mathfrak{U}}$ and the seed ${\bm s}$.    


\subsection{Numerical analysis}

Let $\mu \in\Int_d$ denote the label of the most probable bin of the distribution 
${\cal P}_{\beta}^{({\bm s};\hat{\mathfrak{U}})}$, with the corresponding maximal probability 
denoted by $P^{(0)}({\bm s};\hat{\mathfrak{U}})$ i.e., 
\bea
P({\mathbb B}_\mu|{\bm s};\hat{\mathfrak{U}}) = 
\max_{j\in\Int_d}
\{
P({\mathbb B}_j|{\bm s};\hat{\mathfrak{U}})\} := P^{(0)}({\bm s};\hat{\mathfrak{U}}).
\label{P0:eq}
\eea

\begin{figure}[t]
\includegraphics[scale=0.24]{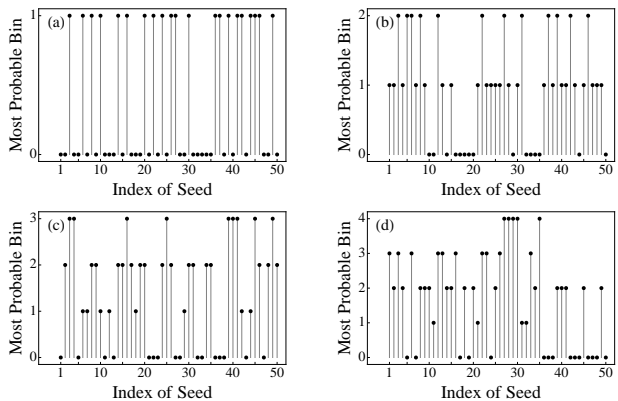}
\caption{
Typical dependence of the label of the most-probable bin $\mu$ on the seed ${\bm s}$, for fixed unitary $\hat{\mathfrak{U}}$.  The vertical gray lines serve as a guide, so that one can easily follow the changes.  The random  $M\times M$ unitary has been chosen according to the Haar measure. The dependence is shown for 2 bins (a); 3 bins (b); 4 bins (c) and 5 bins (d). For the sake of concreteness, the dependence is shown only for the first 50  seeds.  
Other parameters: $M=15$, $N=3$.
}
\label{fig3}
\end{figure}

The distribution ${\cal P}_{\beta}^{({\bm s};\hat{\mathfrak{U}})}$ has been derived by binning all the possible $|{\mathbb S}_{M,N}|$ outcomes in BS  (i.e., all possible photon configurations). 
As is evident from Eq. (\ref{probs:eq}), the probability of the $j$th bin is determined by the probabilities of the individual outcomes that contribute to this bin. Hence, the relative probabilities of the bins in ${\cal P}_{\beta}^{({\bm s};\hat{\mathfrak{U}})}$ are determined by boson-interference effects, as they are manifested in details of the underlying BS distribution $\bar{\cal P}_{{\bm r}}^{({\bm s};\hat{\mathfrak{U}})}$, such as the location of the peaks and dips, relative probabilities of outcomes that contribute to different bins, 
possible correlations between different outcomes, etc. 
To the best of our knowledge, so far the only to some extent relevant results in the literature, are in Refs. \cite{NonUni} and \cite{genBunching}.  More precisely, in Ref. \cite{NonUni} it is shown that BS distribution is far from uniform, which suggests that the structure of the BS distribution is rich, and certainly non-trivial.  In  Ref. \cite{genBunching}, it is shown that boson interference effects can survive binning for bins of sufficiently large size. The results of Ref. \cite{genBunching} , however, are not directly applicable to our work, because they pertain to a particular quantity, namely the probability for all the bosons to appear in the same bin, whereas the focus of the present work is on the most-probable bin of the distribution ${\cal P}_{\beta}^{({\bm s};\hat{\mathfrak U})}$. 

As depicted in Fig. \ref{fig2}, for a fixed Hilbert space, the BS distribution $\bar{\cal P}_{\bm r}^{({\bm s};\hat{\mathfrak U})}$ changes with the seed  and the unitary. Thus, the probabilities of the outcomes that contribute to the different bins  (see Eq. \ref{probs:eq}), also  
vary with both ${\{\bm s};\hat{\mathfrak U}\}$. 
 This dependence is expected to be reflected on the properties of  the derived distribution ${\cal P}_{\beta}^{({\bm s};\hat{\mathfrak U})}$, and thus on the label $\mu$ of the most-probable bin, and the corresponding maximum probability $P^{(0)}$.  
 Both of these quantities are expected to depend on ${\bm s}$ and $\hat{\mathfrak{U}}$.

\begin{figure}[t]
 \includegraphics[scale=0.24]{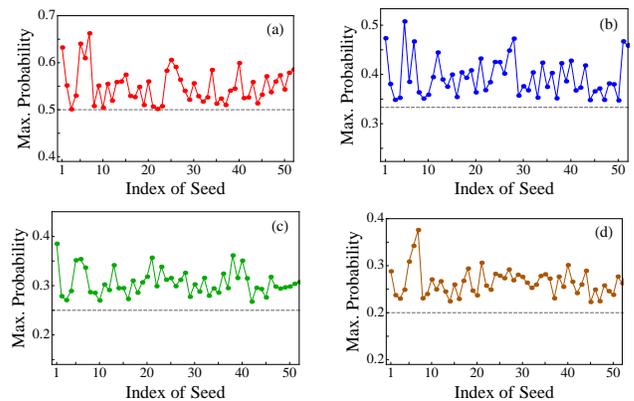}
\caption{(Color online)  
Typical dependence of the maximum probability ${P}^{(0)}({\bm s};\hat{\mathfrak{U}})$ 
on the seed ${\bm s}$, for fixed unitary $\hat{\mathfrak{U}}$. Only the first 50 seeds are shown. 
Other parameters as in Fig. \ref{fig3}.
}
\label{fig4}
\end{figure}

To gain  insight into this dependence  we have performed a series of simulations for various combinations of $\{M,N\}$, and  our main results are presented here.    
As depicted in Fig. \ref{fig3},  for a fixed unitary the bin at which the distribution  ${\cal P}_{\beta}^{({\bm s};\hat{\mathfrak{U}})}$ exhibits its maximum (i.e., the label $\mu$), varies with the seed. The location of the maximum seems to exhibit sudden step-like transitions between the $d$ bins at random and unpredictable seeds. The overall behaviour resembles the random  telegraph signal/noise (also known as burst noise) in electronics, or the quantum jumps in atomic systems (e.g., see chapter 8 of \cite{book_knight}).  
The maximum probability $P^{(0)}({\bm s};\hat{\mathfrak U})$ also varies 
with the seed (see Fig. \ref{fig4}), but as was expected, it always satisfies 
\bea
P^{(0)}({\bm s};\hat{\mathfrak U}) > \frac{1}{d}.
\label{max_cond}
\eea 
The lower bound in this inequality 
refers basically to the case of equally probable bins i.e., to the uniform distribution over $d$ bins.
Hence,  Figs. \ref{fig3} and \ref{fig4} show that for any value of $\mu$ or $P^{(0)}$,  there are many  different seeds that lead to these particular values 
i.e., for any $\{{\bm s};\hat{\mathfrak U}\}$ the function ${\tt MPB}_{{\bm s};\hat{\mathfrak U}}$ is a surjective map, and as such it does not convey much information about the input seed. 

Although these findings are valid for any unitary, the quantitative details (such as the amplitude of the variations in Fig. \ref{fig4}), are expected to depend on the unitary $\hat{\mathfrak U}$. A  key question therefore is the following. Given a randomly chosen 
 unitary $\hat{\mathfrak U}$ and a randomly chosen seed ${\bm s}\in{\mathbb S}_{M,N}$, what  is the {\em a priori} probability ${\rm Pr}(l=\mu|{\bm s};\hat{\mathfrak U})$, for the $l-$th bin  to be the most probable bin of ${\cal P}_{\beta}^{({\bm s};\hat{\mathfrak{U}})}$?  All the seeds are equally probable, and their total number is known. Hence, for any given unitary,  
${\rm Pr}(l=\mu|{\bm s};\hat{\mathfrak U})$ is given by the fraction of seeds for which the corresponding distributions ${\cal P}_{\beta}^{({\bm s};\hat{\mathfrak U})}$ exhibit their maxima at the $l-$th bin. The fraction is expected to vary with $\hat{\mathfrak U}$, hence  we performed simulations for different unitaries chosen at random according to the Haar measure. For each unitary, we calculated the BS distributions for all possible seeds, from which we derived the distributions over $d$ bins, with $2\leq d\ll |{\mathbb S}_{M,N}|$ bins. Hence, we were able to find and keep track of the most-probable bin for each seed and for each unitary.

\begin{figure}[t]
\includegraphics[scale=0.24]{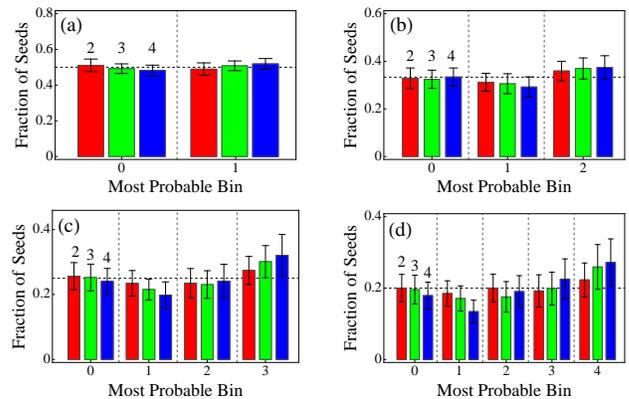}
\caption{(Color online) 
Fraction of seeds for which the distribution ${\cal P}_{\beta}^{({\bm s};\hat{\mathfrak U})}$ exhibits the 
maximum probability at the $l-$th bin. The coloured bars show the fraction averaged over 100 random unitaries chosen according to the Haar measure, whereas the error bars show the standard deviations.
The number of bins is: $d=2$ (a); $d=3$ (b); $d=4$ (c); and $d=5$ (d).  In each case, the vertical dashed lines define the various bins, 
whereas the horizontal dashed lines mark the value $d^{-1}$ corresponding  to equal fractions. 
Bars of different colors refer to three different photon numbers: $N=2$ (red); $N=3$ (green); $N=4$ (blue).  
Other parameters: $M=18$.
}
\label{fig5}
\end{figure}

In Fig. \ref{fig5} we plot the fraction of seeds that result in the most-probable bin ${\mathbb B}_l$ averaged over 100 randomly chosen unitaries, together with the corresponding standard deviations (error bars). For all of the combinations of $\{M,N\}$ we have studied, the average fractions lie within narrow intervals around $1/d$, and the recorded standard deviations reflect the uncertainty due to the choice of the unitary.  Our numerical results suggest that the maximum of ${\cal P}_{\beta}^{({\bm s};\hat{\mathfrak U})}$ may occur at any of the bins, and its actual location depends strongly on the particular seed and unitary under consideration. The bins are not, in general, equally probable images of a given seed ${\bm s}$ under the function ${\tt MPB}_{d; \hat{\mathfrak U}}$, but 
there is no bin which can be considered considerably less probable than the others.

\begin{figure}[t]
\includegraphics[scale=0.24]{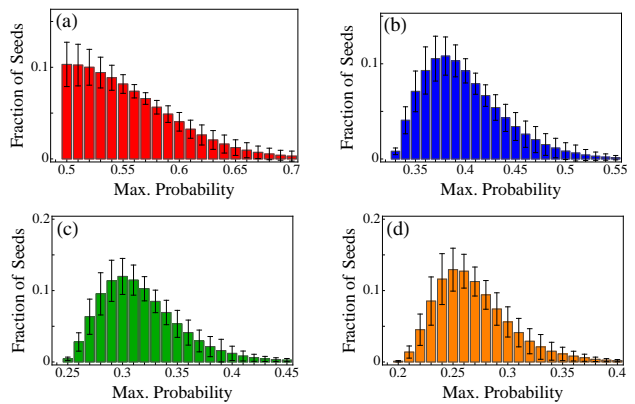}
\caption{(Color online) 
Fraction of seeds for which the maximum probability for the distribution ${\cal P}_{\beta}^{({\bm s};\hat{\mathfrak U})}$ lies between $p$ and $p+dp$. The coloured bars show the fraction averaged over 100 random unitaries chosen according to the Haar measure, whereas the error bars show the standard deviations.  The number of bins is: $d=2$ (a); $d=3$ (b); $d=4$ (c); and $d=5$ (d).  
Other parameters: $M=18$, $N=4$, $dp=0.01$.
}
\label{fig6}
\end{figure}

Besides the label of the most probable bin, another key quantity is the maximum probability of the distribution ${\cal P}_{\beta}^{({\bm s};\hat{\mathfrak U})}$. Our simulations enabled us to quantify the sensitivity of $P^{(0)}({\bm s};\hat{\mathfrak U})$ on both the seed and the unitary. In Fig. \ref{fig6}, we plot the fraction of seeds for which $P^{(0)}$ lies in an interval $[p,p+dp)$, with $d^{-1}<p\leq1$, averaged over 100 random unitaries. As before, the error bars show the standard deviations due to the choice of the unitary.  The recorder standard deviations (error bars) are not very pronounced, and for any $d>2$, the histograms are well approximated by a Gamma distribution.  Moreover we have found that for $d\geq 2$, the histograms become narrower with increasing $d$, while their  mean remains close to $1/d$.

The behaviour depicted in Figs. \ref{fig2}-\ref{fig6}, has been confirmed for various combinations of $\{M,N\}$, within our computational  capabilities (see appendix \ref{app1a}).  
Our results show clearly that the two key parameters of interest (i.e., the label of the most probable bin and the corresponding maximum probability), are rather sensitive to  the seed of the BS, whereas the observed variations with respect to the unitary of the network are not so pronounced.  
These dependences reveal a strong connection between  the distribution ${\cal P}_{\beta}^{({\bm s};\hat{\mathfrak U})}$ of the binned data, and the underlying BS distribution 
$\bar{\cal P}_{{\bm r}}^{({\bm s};\hat{\mathfrak U})}$ from which ${\cal P}_{\beta}^{({\bm s};\hat{\mathfrak U})}$ has been derived. Indeed, as discussed above, the BS distribution depends on both the seed and the unitary, and it is precisely this dependence (at least partially), which is transferred to ${\cal P}_{\beta}^{({\bm s};\hat{\mathfrak U})}$ after the binning. 

\section{A class of problems based on Boson Sampling}
\label{sec4}

Typically, decision problems accept two possible answers {\tt YES} or {\tt NO}, and have the following form: Given a  property ${\tt P}$ and an object ${\tt O}$, decide whether or not the object ${\tt O}$ has the property ${\tt P}$.
Based on the binning of BS data, and in particular on the images of  seeds under ${\tt MPB}_{d;\hat{\mathfrak U}}$, we can define the following class of decision problems. 

\begin{definition} 
\label{def2}  Class of BS decision problems {\rm (${\tt BSDP}_f$)}. \\
Parameters: number of photons $N>2$, 
$M\times M$ unitary $\hat{\mathfrak U}$, number of bins 
$2\leq d\ll |{\mathbb S}_{M,N}|$.\\
Input: $n$ different seeds chosen at random and independently from a uniform distribution over ${\mathbb S}_{M,N}$, with $n\ll |{\mathbb S}_{M,N}|$. If necessary 
a set of ancillary integers ${\bm y}:=(y_1,y_2,\ldots).$ \\ 
Query: Decide whether a function $f({\bm x},{\bm y})$, 
with ${\bm x}:=(x_1,x_2,\ldots,x_n)$ and 
$x_j = {\tt MPB}_{d;\hat{\mathfrak U}}({\bm s}_j)$,
satisfies a property ${\tt P}$ or not. 
\end{definition}

Cryptographic applications are usually built on  function problems, which are closely related to decision problems,  but the possible answers are more complex than a simple {\tt YES} or {\tt NO}.  A class of function problems that are 
associated with ${\tt BSDP}_f$, is the following.

\begin{definition}  
\label{def3} Class of BS function problems ${\tt BSFP}_f$.\\
Parameters: Same as in ${\tt BSDP}_f$. \\
Input: Same as in ${\tt BSDP}_f$.\\ 
Query: Calculate $f({\bm x},{\bm y})$.
\end{definition}

Each problem in ${\tt BSDP}_f$ or ${\tt BSFP}_f$ is  essentially parametrized by the mathematical function $f$.  
Moreover, as will be discussed in Sec. \ref{sec4c},
in order for a problem to be useful in practise, we require that the number of images associated with it be considerably smaller than the total number of seeds i.e.,  
$n\ll |{\mathbb S}_{M,N}|$. 

\begin{example} The class ${\tt BSDP}_f$ includes, but is not limited to, decision problems such as: 
Is $\max\{{\bm x}\}$ or $\min\{{\bm x}\}$ equal to 
a given integer $y_1$?  Is the sum $X:=\sum_{j=1}^n x_j$ larger than $y_1$? Is 
$X$ an even number?  Is the integer $y_2$ the greatest common divisor ({\tt gcd}) of $X$ and $y_1$? 
The class ${\tt BSFP}_f$ includes, but is not limited to, problems such as: finding the $\max\{{\bm x}\}$ or $\min\{{\bm x}\}$,  estimation of the sum $X$, and  estimation of ${\tt gcd}(X, y_1)$. 
\end{example}

It is worth recalling here that $x_j$ is the label of the most-probable bin when the data of  session 
${\tt BS}({\bm s};\hat{\mathfrak U})$ are binned with respect to $d$ disjoint bins. Each bin refers to a subset of all the possible 
outcomes written in their integer representation.  Hence, instead of defining  decision/function problems with respect to the labels 
of the most-probable bins in $n$ BS sessions, one may also define analogous problems with respect to specific outcomes in the 
most-probable bins ${\mathbb B}_{x_1}$, ${\mathbb B}_{x_2}$, etc. 

\begin{example}
One may ask  whether the $i$th outcome in bin ${\mathbb B}_{x_j}$  is larger than some integer $y_1$. 
Or what is the $i$th outcome in the bin ${\mathbb B}_{x_j}$, where $x_j$ is the $j$th largest integer in ${\bm x}$? 
\end{example}

In general, one may think of many different decision and function problems, within the present theoretical framework. Whether some of these problems are more promising than others can be decided only on the basis of specific applications, after a thorough security analysis. Such an investigation goes beyond the scope of the present work, which aims at proposing a general framework for the definition of BS-based decision and function problems, whose solution may be  computationally hard. One main feature of the aforementioned examples is that their solution requires the estimation of the images $x_1, x_2, \ldots, x_n$ for  each one of the given randomly chosen 
seeds ${\bm s}_1, {\bm s}_2, \ldots, {\bm s}_n$. 
In the following subsections we discuss different approaches to such an estimation, in order to gain some insight into the computational complexity of 
${\tt BSDP}_f$ and ${\tt BSFP}_f$.


\subsection{Brute-force numerical solution}
\label{sec4a}

Assuming that one can perform computations of arbitrary accuracy, 
the key question for the solution of problems in ${\tt BSDP}_f$ and ${\tt BSFP}_f$,  is about the optimal algorithm for the 
estimation of the image  of a randomly chosen seed under ${\tt MPB}_{d;\hat{\mathfrak U}}$. In the absence of any relevant results,  
we will discuss a brute-force numerical approach to the estimation of 
${\tt MPB}_{d;\hat{\mathfrak U}}({\bm s})$ for a particular session  ${\tt BS}({\bm s};\hat{\mathfrak U})$,  which 
relies on the estimation of the probabilities of all the $d$ bins $\{P({\mathbb B}_j|{\bm s};\hat{\mathfrak U})\}$, by means of Ryser's algorithm.  
The image of ${\bm s}$ under 
 ${\tt MPB}_{d;\hat{\mathfrak U}}$ is the label of the most probable bin and can be  readily obtained from  $\max_j\{P({\mathbb B}_j|{\bm s};\hat{\mathfrak U})\}$.  In view of Eq. (\ref{probs:eq}), the  evaluation of the probability for the $j$th bin $P({\mathbb B}_j|{\bm s};\hat{\mathfrak U})$
requires the evaluation of at least $|{\mathbb B}_j|$ different probabilities 
$P({\bm r}|{\bm s};\hat{\mathfrak U})$, which pertain to the underlying BS distribution 
$\bar{\cal P}_{{\bm r}}^{({\bm s};\hat{\mathfrak U})}$. Using Eqs.  (\ref{w:eq}) and 
(\ref{w2:eq}), for $d\ll |{\mathbb S}_{M,N}|$ we have, 
$|{\mathbb B}_j| \simeq |{\mathbb S}_{M,N}| d^{-1}$, and thus  in the dilute limit of BS 
\[
|{\mathbb B}_j| \simeq \frac{1}{d}\binom{M}{N}.
\]
For $M\sim N^2$ and $N\gg 1$, one can readily show that 
\bea
|{\mathbb B}_j| > \binom{M}{N-1} \sim \left (\frac{M}{N} \right )^{N}. 
\label{bin_size2:eq}
\eea
 
As discussed in Sec. \ref{sec2}, each one of the conditional probabilities $P({\bm r}|{\bm s};\hat{\mathfrak U})$,  requires the estimation of a permanent of a complex $N\times N$ matrix \cite{complex1,BSintro,complex2,complex3,valiant, Scheel}. 
So far,  the most efficient known algorithm for the estimation of permanents is Ryser's algorithm, which for 
a single $N\times N$ matrix uses $N 2^{N}$ floating-point arithmetic operations  \cite{ryser}.  
This means that the calculation of $\xi$ permanents  on a computer that performs  
$\nu$ floating-point arithmetic operations per second (flops), will take time at least 
\bea
T_{\rm ryser} = \xi \frac{ N 2^{N} }{\nu}\, {\rm sec.}
\label{cpu_time_id}
\eea

According to Eq. (\ref{bin_size2:eq}), for $N = 13$ and $M = 100$, the calculation of  
$P({\mathbb B}_j|{\bm s};\hat{\mathfrak U})$ requires the 
estimation of about $|{\mathbb B}_j|> 10^{15}$ probabilities 
$P({\bm r}|{\bm s};\hat{\mathfrak{U}})$, and thus the same number of permanents. 
Equation  (\ref{cpu_time_id}) suggests that such a task on a (single-core) processor that 
has theoretical performance $10^{12}$ flops, will take at least $10^{8}$ sec, i.e., more than 10 years.  
Given that the calculation of each permanent is independent of the others, 
this time can be reduced considerably, by distributing the task  
in many processors.  For instance, the task could be performed within hours, on a supercomputer  
with about  $3\times 10^4$ such processors, where each processor is burdened with the calculation of about 
$3\times 10^{10}$ permanents. To the best of our knowledge, however, there are only a few 
supercomputers with such characteristics world-wide \cite{supercomputers}. For larger values of $N$ and/or 
$M$, the number of permanents 
required for the calculation of $P({\mathbb B}_j|{\bm s};\hat{\mathfrak{U}})$  increases dramatically, 
and the task becomes practically impossible for current supercomputers.  
For instance, when $N=16$ and $M = 200$,   $|{\mathbb B}_j|>10^{22}$  
and the task on a supercomputer with about $10^6$ CPU cores at $10^{12}$ flops,  
would take more than $10^{10}$sec i.e., more than $10^3$ years. 

Of course, these are theoretical estimates for the evaluation of the probability for a single bin in a particular session ${\tt BS}({{\bm s};\hat{\mathfrak U}})$. 
In practise one has to expect longer times, 
due to imperfect performance of processors,  poor implementation of Ryser's algorithm, 
and additional unavoidable intermediate steps in the computation, such as memory allocation, execution of 
input/output commands, etc. 
The estimates have been also confirmed by our simulations, and provide strong evidence  that the brute-force estimation of the images of $n$ randomly 
chosen seeds on a classical computer  becomes very quickly a hard problem for a small number of bins, as we increase the  photon number $N\gg 1$ in a network of $M\sim N^2$ modes. 
This is because the solution involves the calculation of permanents by means of Ryser's algorithm, for  about 
$(MN^{-1})^N$ complex matrices of dimensions $N\times N$ (see also additional details in appendix \ref{app1}). 


\subsection{Solution by non-boson samplers}
\label{sec4b}

In view of the results of the previous subsection, the natural question arises is whether the image of an $N$-boson seed under ${\tt MPB}_{d;\hat{\mathfrak U}}$ 
can be evaluated in the framework of non-bosonic particles such as fermions or 
distinguishable particles.  In the case of fermions,  the transition probability 
$P({\bm r}|{\bm s};\hat{\mathfrak U})$  is related to the determinant of the $N\times N$ 
matrix  $\hat{\mathfrak U}_{{\bm s},{\bm r}}$ (see Sec. \ref{sec2}), whereas in the case of distinguishable particles the transition probability pertains to a matrix with real elements i.e., 
$P({\bm r}|{\bm s};\hat{\mathfrak U})\propto {\rm Per}[|\hat{\mathfrak U}_{{\bm s},{\bm r}}|^2]$ \cite{TichyPRL14,NonBS,NonBS2}. 
As a result, and by contrast to BS, non-boson sampling can be simulated efficiently on classical computers.

Let $x^{(\iota)}$ denote the image of the $N-$particle seed ${\bm s}$ under ${\tt MPB}_{d;\hat{\mathfrak U}}$, when particles of type $\iota$ are used. 
The label $\iota$ can be: B (for bosons), F (for fermions) or D (for distinguishable particles). Given that ${\tt MPB}_{d;\hat{\mathfrak U}}({\bm s})$ pertains to the binned data of the session ${\tt BS}({\bm s};\hat{\mathfrak U})$, the question basically is whether boson-interference effects survive the binning,  so that $x^{({\rm B})} \neq x^{({\rm F/D})}$.  As shown in Ref. \cite{genBunching}, one can distinguish between bosons and non-bosons by looking at the probability for all of the bosons to occupy the same bin. By contrast, here we focus on another quantity, i.e., the image of the seed ${\bm s}$ under ${\tt MPB}_{d;\hat{\mathfrak U}}$, and we wish to investigate whether interference effects survive the binning, with respect to this particular quantity.  

To address this question, we evaluated $x^{({\rm B})}$,  $x^{({\rm D})}$ and  $x^{({\rm F})}$, for different randomly chosen unitaries,  and for all the possible  $N-$particle seeds.  In this way, for each combination of $\{d,\hat{\mathfrak U}\}$, we were able to estimate the average fraction of seeds for which $x^{({\rm B})} = x^{({\rm D})}$ and $x^{({\rm B})} = x^{({\rm F})}$, as well as the standard deviations with respect to the choice of the unitary. According to definitions \ref{def2}  and \ref{def3},  the seeds are chosen at random from a uniform distribution over the set of all possible seeds ${\mathbb S}_{M,N}$.  Hence, 
the average fraction of seeds we have estimated quantifies the probability of collision $p_{\rm col}$ i.e.,  the probability for a randomly chosen bosonic seed to have the same image as a non-bosonic seed,  under the same ${\tt  MPB}_{d;\hat{\mathfrak U}}$.

\begin{figure}
\includegraphics[scale=0.9]{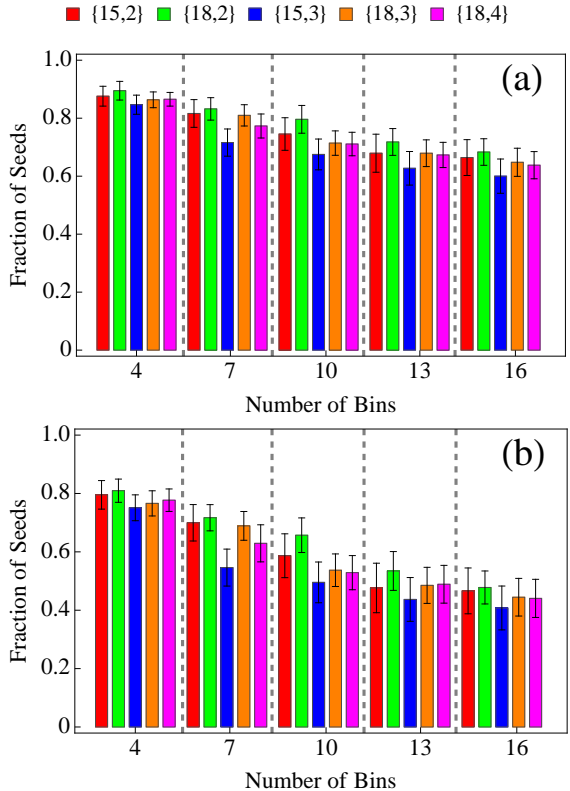}
\caption{ 
(Color online) 
Fraction of seeds for which  for which $x^{({\rm B})} = x^{({\rm D})}$ (a) and $x^{({\rm B})} = x^{({\rm F})}$ (b), for various combinations of $\{M,N\}$ and different numbers of bins.  The coloured bars show the fraction averaged over 100 random unitaries chosen according to the Haar measure, whereas the error bars show the standard deviations. Bars of different colors refer to different 
combinations of parameters $\{M,N\}$, given on the top of the figure.
}
\label{fig7}
\end{figure} 
 
As depicted in Fig. \ref{fig7},  for $d<5$ the probability of collision is $p_{\rm col}\simeq 1$. 
For such small values of $d$ therefore, ${\tt  MPB}_{d;\hat{\mathfrak U}}$ becomes insensitive to the underlying 
statistics, and  the image of any bosonic seed coincides with the image of 
the corresponding non-bosonic seed with high probability. As we increase $d$, however, the probability of collisions decreases, which means that boson-interference effects survive the binning and are reflected in the sensitivity of ${\tt  MPB}_{d;\hat{\mathfrak U}}$ on the type of particles in the input. This fact has been confirmed for various sizes of the Hilbert 
space (i.e., for various combinations of $\{M,N\}$). Although we were not able to perform simulations for systems of  complexity higher than 
$\{M=18,N=4\}$, the behaviour depicted in Fig. \ref{fig7} suggests clearly that 
for fixed $d$, the variation of $p_{\rm col}$ with the size of the Hilbert space are not so pronounced. Hence, one can achieve $p_{\rm col}< 1$ for a number of bins  that is considerably smaller than the size of the Hilbert space. For instance, in the 
case of distinguishable particles and for $d=16$ bins,  
we find $p_{\rm col}\simeq 0.7$ for  sizes $|{\mathbb S}_{M,N}|$ ranging from  about $100$ to about $3000$. 

The key point now is that these estimates pertain to a single randomly  chosen seed, whereas by definition, problems in ${\tt BSDP}_f$ and ${\tt BSFP}_f$ pertain to the images of $n$ random independently  chosen seeds. The estimation of all the $n$ images with high probability is a necessary condition for a non-boson sampler to be considered as capable of solving problems in ${\tt BSDP}_f$ and ${\tt BSFP}_f$ efficiently. 
The probability for a non-boson sampler to succeed in estimating correctly all of the $n$ images is $(p_{\rm col})^{n}$ i.e., it decreases exponentially with $n$. 
For instance, assuming $p_{\rm col}\simeq 0.8$, we have $(p_{\rm col})^{n}\simeq 0.3$ for $n=5$. So, for  sufficiently large values of $n\ll |{\mathbb S}_{M,N}|$, non-boson samplers are not expected to be capable of solving efficiently any problem in ${\tt BSDP}_f$ or ${\tt BSFP}_f$.  One has to focus, however, on problems whose solution depends strongly on the sequence of the $n$ images, and some examples are given above. 


\subsection{Solution by linear optics}
\label{sec4c}

As discussed in the previous subsections, neither non-boson samplers or a brute-force numerical approach are capable of solving judiciously  designed problems in classes ${\tt BSDP}_f$ and ${\tt BSFP}_f$. In order, however, for such problems to be useful in practise, there should exist an approach which allows for their  efficient solution. 

Contrary to classical computers, it has been shown that  BS devices that rely on linear optics  can produce samples from a given distribution efficiently. 
Equation (\ref{probs:eq}) shows that in the framework of experimental BS,  
the probability for the $j$th bin $P({\mathbb B}_j|{\bm s};\hat{\mathfrak U})$ 
can be calculated by post-processing the data obtained during the experimental runs. 
As discussed in Sec. \ref{sec2}, when sampling from the actual BS distribution, each run of the experiment is expected to yield a ``click" for one of the $|{\mathbb S}_{M,N}|$ possible outcomes, 
which can be subsequently assigned to one of the  $d$  possible bins.  
Analysing the frequency of occurrences of the bins in ${\mathscr N}$ experimental runs for fixed $\{{\bm s},\hat{\mathfrak{U}}\}$, one will obtain information about the probabilities $\{P({\mathbb B}_j|{\bm s};\hat{\mathfrak U})\}$ associated with all of the bins.  For the estimation of the image of a randomly chosen seed ${\bm s}$ under 
${\tt MPB}_{d;\hat{\mathfrak U}}$,  one does not have to reconstruct the entire distribution i.e. to obtain 
faithful estimates for all of the probabilities $\{P({\mathbb B}_j|{\bm s};\hat{\mathfrak U})\}$. 
Instead, to estimate the image  ${\tt MPB}_{d;\hat{\mathfrak U}}({\bm s})$, it is sufficient to faithfully estimate the 
 largest probability in the distribution ${\cal P}_{\beta}^{({\bm s};\hat{\mathfrak U})}$,  given by Eq. (\ref{P0:eq}).
Our task in this subsection is to estimate the required sample sizes to this end. 

In practise, due to imperfections in the implementation and statistical deviations in finite samples, one essentially will be able to approximate the maximum probability $P^{(0)}$ only to some accuracy.  
Faithful estimation of $P^{(0)}$ means that one can ensure 
\bea
{\rm Pr}[|{Q}^{(0)}({\bm s};\hat{\tilde{\mathfrak U}}) - P^{(0)}({\bm s};\hat{\mathfrak U}) |< \varepsilon]>1-\eta, 
\label{Chernoff_Ptot}
\eea
for some  accuracy $\varepsilon$ and confidence $1-\eta$, with $\varepsilon,\eta \ll 1$.  
The empirical  probability  ${Q}^{(0)}({\bm s};\hat{\tilde{\mathfrak U}}) $, is the frequency of occurrence of the most probable bin in the ${\mathscr N}$ experimental runs for seed ${\bm s}$. In order to make clear the effects of imperfections, we have introduced the implemented unitary $\hat{\tilde{\mathfrak{U}}}$, which is assumed to be different from 
(but close to) the theoretically expected unitary  $\hat{\mathfrak{U}}$.

For a given BS set-up, the contributions of statistical deviations and  
imperfections  can be 
separated by employing  the triangle inequality 
\[
|Q^{(0)} - P^{(0)}|\leq |Q^{(0)}-R^{(0)}|+|R^{(0)}-P^{(0)}|.
\] 
Here, ${R}^{(0)}({\bm s};\hat{\tilde{\mathfrak U}}) := \max_{j}\{R({\mathbb B}_j|{\bm s};\hat{\tilde{\mathfrak U}})\}$
denotes the approximate probability in the presence of the imperfections, if infinitely large samples were possible. In general, the lowest attainable error and the highest possible confidence in Eq. (\ref{Chernoff_Ptot}) are  determined by the imperfections, whose control is a necessary precondition for  BS to be of some usefulness (even as a paradigm of quantum supremacy). Related work has revealed that BS is rather robust to various types of imperfections \cite{error1,error2,error2b,error3,error4}. 
Assume that one can ensure an overall accuracy $\delta<\varepsilon$  
in the implementation with uncertainty at most $\gamma<\eta$ i.e., 
\[
{\rm Pr}[|R^{(0)}-P^{(0)}| < \delta] > 1-\gamma.
\]
Then, 
in view of the triangle inequality, to fulfil condition (\ref{Chernoff_Ptot}) for the given $\{\delta,\gamma\}$, one needs to  suppress statistical deviations so that 
\bea
{\rm Pr}[|Q^{(0)} -R^{(0)} |<\bar{\varepsilon}]  >1-x, 
\label{Chernoff_Sample}
\eea
for 
\[x < \frac{\eta-\gamma}{1-\gamma}\,
\textrm{ and } \, 
\bar{\varepsilon}:=\varepsilon-\delta,\]
where we have used the independence of deviations due to imperfections and to finite sampling. 
The crucial question now is what  are the sample sizes that can ensure condition (\ref{Chernoff_Sample}). 

Instead of bounding the absolute error as dictated by Eq. (\ref{Chernoff_Sample}), we will bound the relative error which is 
a  stricter requirement because $\bar{\varepsilon}R^{(0)}  \leq \bar{\varepsilon}$.
Employing the Chernoff bound \cite{C-H-cite,book2,SSS,book3} we have 
\begin{widetext}
\bea
 {\rm Pr}\left [ |Q^{(0)} ({\bm s};\hat{\tilde{\mathfrak U}})-R^{(0)} ({\bm s};\hat{\tilde{\mathfrak U}}) |\geq \bar{\varepsilon} R^{(0)}({\bm s};\hat{\tilde{\mathfrak U}})  \right ] 
 \leq
 2\exp\left [
 -\frac{\bar{\varepsilon}^2 R^{(0)} ({\bm s};\hat{\tilde{\mathfrak U}})  {\mathscr N} }{3}
 \right ].
 \nonumber 
\eea
\end{widetext}
In view of condition (\ref{max_cond}), for any $\{d,{\bm s},\hat{\tilde{\mathfrak{U}}}\}$, the maximum probability will 
be always larger than  $d^{-1}$. Hence, we have 
\bea
{\rm Pr}\left [ |Q^{(0)}-R^{(0)}  |\geq \bar{\varepsilon} R^{(0)}  \right ] 
 \leq  2\exp\left [
 -\frac{\bar{\varepsilon}^2  {\mathscr N}  }{3d}
 \right ],
\eea
which has to be smaller than $x$ so that condition  (\ref{Chernoff_Sample}) is satisfied. 
Solving for  ${\mathscr N}$ we obtain that one needs a sample size of  at least 
\bea
{\mathscr N}_{\min}(d,{\bm v}) &=& \frac{3 d}{(\varepsilon-\delta)^2}\ln\left [\frac{2(1-\gamma)}{\eta-\gamma}\right ],
\label{SampleSize_min}
\eea 
where ${\bm v}:=\{\varepsilon, \delta, \eta,  \gamma\}$.

Equation (\ref{SampleSize_min}) is independent of the BS parameters, and depends only on the number of bins, as well as on the accuracy and confidence levels, with the former being far more costly than the latter. 
This means that for any BS session, with given partition of ${\mathbb S}_{M,N}$ into $d$ bins, 
the most-probable bin can be found with a finite number of measurements, which is independent of the 
size of the Hilbert space.  For the sake of comparison, it is worth noting here that 
the estimation of the most-probable outcome in the actual BS distribution $\bar{\cal P}_{\bm r}^{({\bm s};\hat{\mathfrak U})}$,  
would require an exponentially large number of measurements $(\sim |{\mathbb S}_{M,N}|^{0.7})$ (see appendix \ref{app1b}). This fact reveals the pivotal role of binning in the definition of mathematical decision/function problems that can be solved efficiently by passive linear interferometers. Moreover, it shows that a decision/function problem in classes ${\tt BSDP}_f$ and  ${\tt BSFP}_f$ can be practical only if the total number of images involved is much smaller than $|{\mathbb S}_{M,N}|$. 

Given that the worst-case scenario has been adopted at various stages of the above derivation, our estimation (\ref{SampleSize_min}) is  expected to be rather pessimistic. 
In practise, smaller samples are expected to be required, especially if one employees  
sophisticated sampling techniques, which rely on the gradual increase of the sample size (e.g., sequential sampling \cite{SampleBooks,SampleSize}). 
In any case, it has to be emphasized that the faithful estimation of the largest probability by means of sampling does not eliminate the possibility of failure. More precisely, even in the case where the  obtained  sample size is  sufficiently large to ensure Eq. (\ref{Chernoff_Ptot}), there is always a small probability for the estimation to fail, which is represented by the uncertainty $\eta$. Such events stem from the fact that in practise, as a result of finite samples and imperfections, the achievable confidence levels will be always smaller than 1. 

\begin{figure}
\includegraphics[scale=0.9]{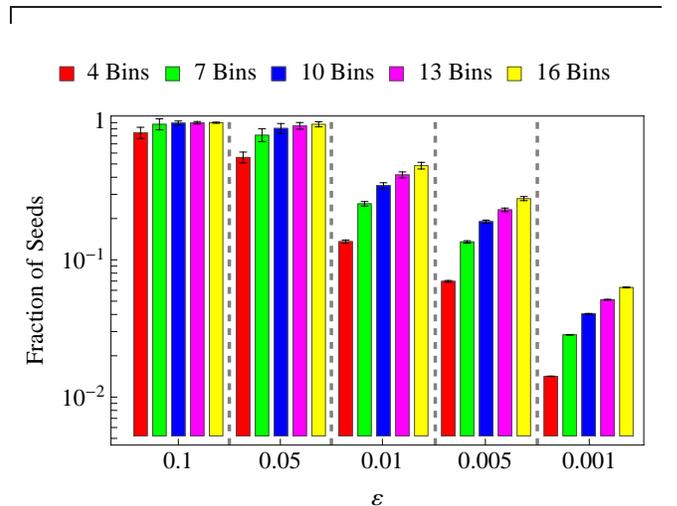}
\caption{ 
(Color online) Fraction of seeds for which the two largest probabilities in the distribution 
${\cal P}_{\beta}^{({\bm s};\hat{\mathfrak U})}$ differ by less than $\varepsilon$, for various numbers of bins. The coloured bars show the fraction averaged over 100 random unitaries chosen according to the Haar measure, whereas the error bars show the standard deviations.
Parameters: $M=18$, $N=4$.
}
\label{fig8}
\end{figure}

The estimation of the largest probability $P^{(0)} $, however, may also fail even if one were able to achieve absolute confidence (i.e.,  $\eta=0$).  This may happen when the accuracy provided by the obtained 
sample size is not sufficiently large to distinguish between the two largest probabilities $P^{(0)} $ and $P^{(1)}$ 
in ${\cal P}_{\beta}^{({\bm s};\hat{\mathfrak U})}$; that is, when $P^{(0)} -P^{(1)} \leq \varepsilon$.  Unfortunately, besides the accuracy $\varepsilon$, the presence of such ambiguities also depends on the structure of the BS  distribution from which ${\cal P}_\beta^{({\bm s};\hat{\mathfrak U})}$ is derived,  and thus on all of  the parameters $\{d,{\bm s},\hat{\mathfrak{U}}\}$. Given that for sufficiently large values of $\{M,N\}$, the BS distribution is unknown (i.e., it cannot be calculated numerically),  one cannot know in advance whether such an ambiguous scenario is pertinent to a particular BS session or not. 
To obtain an estimate for the fraction of seeds for which  $P^{(0)} -P^{(1)} \leq \varepsilon$, we 
have resorted to numerical simulations for various randomly chosen unitaries, and our main results are 
summarized in Fig. \ref{fig8}. First of all, it is worth noting that the uncertainty due to the choice of the unitary $\hat{\mathfrak{U}}$ 
is negligible. Secondly, for fixed $\varepsilon$,  the fraction of seeds for which 
$P^{(0)} -P^{(1)} \leq\varepsilon$ increases with increasing number of bins. As long as, however, we focus on $\varepsilon\lesssim  5\times 10^{-2}$, 
it is only about $10\%$ of the seeds or less that may lead to ambiguities.  The behaviour depicted in  Fig. \ref{fig8} refers to $M=18$ and $N=4$, and it is very close to the behaviour we have found for other combinations of $\{M,N\}$. 


\section{Concluding Remarks}  
\label{sec5}

We have defined a rather general theoretical framework for the design of BS-based decision and functions problems. Such problems,  if they are computationally hard,  may pave the way to applications in modern cryptography.   The proposed framework  pertains to the binning of data  from BS experiments, and the 
quest for the  most-probable bin. 

Experimental BS typically suffers from different sources of errors and losses \cite{error1,error2,error2b,error3,error4}.  
BS devices, however, are far less demanding than full-power quantum computers, as they 
require only passive linear optics, single-photon sources, and standard photo-detection systems. 
Linear-optics elements are rather accurate allowing for reliable implementation of 
a given unitary \cite{ExpBS1,ExpBS2,ExpBS3,ExpBS4,ExpU}, while one can  
estimate an upper bound on the variation distance of the approximated distribution from 
the ideal \cite{error1}.  
In any case, it has to be  emphasized here, that the computational complexity  of  BS in the presence of 
imperfections is not so clear for now, but there is strong evidence that  when the approximate  distribution is close to the ideal one and for low losses, the problem is still computationally hard \cite{complex1, error2,error2b,error3}. 

Throughout this work we have assumed that the number of modes scales quadratically with
the number of photons, which essentially guarantees that at the output of the network the photons occupy different modes \cite{complex1}. This assumption does not in any case affect the main ideas presented in this work, 
which are also applicable if the number of modes scales linearly with the number of photons. 
In this case, however, the hardness of BS has not been shown rigorously, although it is very plausible \cite{complex1}.  
Furthermore, from the experimental point of view, one would need photo-resolving detection systems at the output, so that to distinguish between different possible outcomes. 

In closing, we would to emphasize once more that the purpose of this work was to set a general theoretical framework for BS-based mathematical problems, which may have practical applications that go beyond the 
proof of quantum supremacy. The general framework we presented opens up new directions of research in quantum technologies, because it is rather flexible and allows for the definition of various mathematical problems.  
Our extensive numerical simulations and estimates provide strong evidence on the  computational hardness of the proposed classes of decision/function problems, but a  rigorous proof remains an open question. Moreover, whether specific problems are more promising than others can be decided only on the basis of specific applications, and after the appropriate security analysis.

\section {Acknowledgements} The authors acknowledge financial support from UK EPSRC grant EP/I01245/1. GMN is grateful to S. M. Barnett and T. Brougham for their support and hospitality at the University of Glasgow, where the main part of this work was performed. We thank S. M. Barnett and P. Lambropoulos for helpful comments and suggestions.


\appendix

\begin{table}
\centering
\caption{Time complexity of BS.  
The input photon configuration ${\bm s}$ is fixed, and  the estimation of the conditional probabilities  
$P({\bm r}|{\bm s};\hat{\mathfrak U})$ for all of the $\binom{M}{N}$ different output configurations ${\bm r}$, is achieved by means of Ryser's algorithm \cite{ryser}. }
\label{tab1}
\begin{ruledtabular}
\begin{tabular}{c|c|c|c|c}
$M$ & $N$ & configurations & \multicolumn{2}{c}{CPU Time (ms)}             \\ 
\hline
     &  &   & Node x1$^{{\rm a}}$  & Node x10$^{{\rm b}}$ \\ \hline
4  & 2 & 6  & 3  & 2\\ 
8  & 2 & 6  & 4  & 3\\ 
16  & 2 & 6  & 9  & 4\\ \hline
9  & 3 & 84  & 7  & 5\\ 
18  & 3 & 816  & 58  & 20\\ 
24  & 3 & 2024  & 157  & 54\\ \hline
16  & 4 & 1820  & 192  & 70\\ 
20  & 4 & 4845  & 554  & 184\\ \hline
25  & 5 & 53130 & 5059  & 4110\\ \hline
30  & 6 & 593775 & 115594  & 101540\\ 
\end{tabular}
\end{ruledtabular}
\footnotetext[1]{Node x1: Intel(R) Xeon(R) CPU E5-2650 v2 @ 2.60GHz} 
\footnotetext[2]{Node x10: Intel(R) Xeon(R) CPU E5-2640 v3 @ 2.60GHz}
\end{table}

\section{Simulations on Boson Sampling}
\label{app1}

Throughout our simulations the random complex unitary $\hat{\mathfrak U}$ was  chosen randomly according to the Haar measure. For reasons discussed in the main sections of the 
paper, we focused on the dilute limit of BS corresponding to $M\sim N^2$ modes, where the probability of observing configurations with multiple occupancy of modes is small \cite{complex1,BSintro}, and the problem can be analysed accurately  in the framework of the $\binom{M}{N}$ different ``collision-free" configurations. 

\begin{figure}[t]
\includegraphics[scale=0.45]{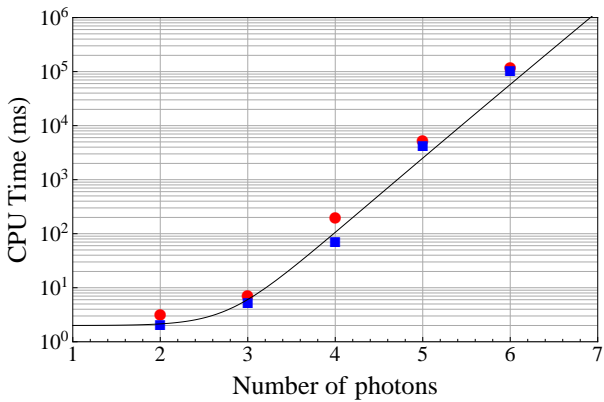}
\caption{(Color online) As in table \ref{tab1}. The symbols pertain to the data  for $M=N^2$ and with $N=2,3,4,5, 6$. 
Circles and squares refer to node x1 and node x10 respectively.  
The solid line is a fit to the data according to the law of Eq. (\ref{cpu_time}).}
\label{fig-A1}
\end{figure}

\subsection{Performance}
\label{app1a}
The simulations were performed on two different nodes of the cluster at FORTH, 
and some of the recorded CPU times are given in table \ref{tab1}. Note that the depicted CPU times refer to the 
estimation of all of the $\binom{M}{N}$ different probabilities $P({\bm r}|{\bm s};\hat{\mathfrak{U}})$ 
for given seed and unitary. 
The nodes of the cluster are different generations of the same processor (one is supposed to be faster than the other), and for both of them we observe a rapid increase of the CPU time as we increase the complexity of the problem (i.e., for increasing $M,N$).  This is a global behaviour we have observed,  and in order to obtain a  law on the performance of our code, we tried to fit our data with a curve of the form 
\bea
T_{\rm cpu} = A\times N \times 2^{B\times N}+C,
\label{cpu_time}
\eea
which is in agreement with the time-complexity of  Ryser's algorithm (see Sec. \ref{sec4a}).
 As shown in Fig. \ref{fig-A1}, our data are 
well approximated by such a curve with parameters $A \simeq 1.8915\times 10^{-4}$ ms, $B \simeq 4.267$ 
and $C\simeq 2.36$ ms. 
As was expected, the recorded CPU times depicted in  table \ref{tab1} and in Fig. \ref{fig-A1} are well above the 
theoretical estimate (\ref{cpu_time_id}). This is mainly because  Eq. (\ref{cpu_time_id}) refers only to 
the performance of Ryser's algorithm for a given number of  $N\times N$ matrices, whereas 
our program performs many additional tasks such as the generation of the random unitary, 
the storage of data at various stages, execution of input/output commands, etc.  The recorded growth 
of the CPU time with $N$, however, does follow the exponential law expected for Ryser's algorithm.

It is worth emphasizing here that according to our simulations the CPU time depends strongly 
on the total number of outcomes $|{\mathbb S}_{M,N}|$ (i.e., the size of the Hilbert space). 
This is not surprising, because as discussed in Sec. \ref{sec4a},    
the total number of permanents involved in the calculation of 
a BS distribution is proportional to the size of the Hilbert space, and  the same 
holds for the total number of required floating-point arithmetic operations. 
As a result, for moderate values of $N$ where the calculation of the 
permanent of a singe $N\times N$ matrix is possible, 
the estimation of the entire BS distribution can be a hard task, because  it involves the calculation 
of $|{\mathbb S}_{M,N}|\sim (MN^{-1})^N$ different permanents. 
Consider for instance the case of $N=13$ and $M=N^2$. 
The calculation of the permanent of a singe $13\times 13$ matrix
requires about $10^5$ operations, and thus it is an easy task for a processor that 
performs about $10^{12}$ flops. 
The calculation of $|{\mathbb S}_{M,N}|\sim N^N$ different permanents, however, requires $10^{20}$ 
operations, and it would take more than 10 years on the same processor. In order to achieve the same number 
of operations for a single matrix, and thus the same level of computational hardness, 
one needs more than 40 photons.  

All of these estimates (as well as those of Sec. \ref{sec4a}) hold for computations that rely on Ryser's algorithm, and for input states that are pure photon-number states. 
For such states it is known that the BS distribution is associated with the calculation of permanents of complex matrices, which is a $\#$P-hard problem for classical computers. If one considers, for instance, coherent states at the input, then the calculation requires only matrix multiplications, and the problem of BS can be simulated efficiently on classical computer irrespective of the 
size of the Hilbert space \cite{complex1}.

\section{Maximum probability of BS distribution}
\label{app1b}
\begin{figure}[b]
\includegraphics[scale=0.95]{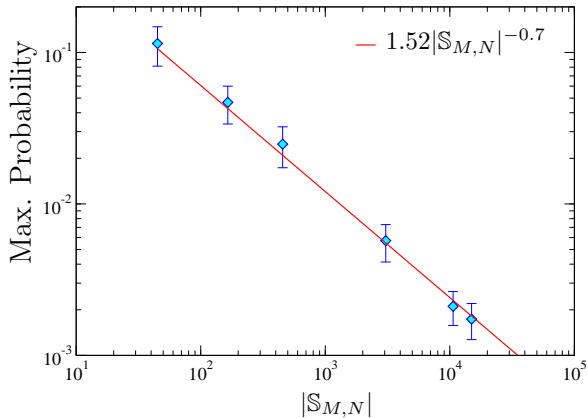}
\caption{(Color online) The maximum probability in the BS distribution $\bar{\cal P}_{\bm r}^{({\bm s};\hat{\mathfrak U})}$,  as a function of the size of the Hilbert space (number of outcomes). The data pertain to various 
combinations of $\{N,M\}$, with $M\geq N^2$, while $|{\mathbb S}_{M,N}| = \binom{M}{N}$. The error bars show the standard 
deviations with respect to the seeds.}
\label{fig-A2}
\end{figure}

Our simulations enabled us to analyse various aspects of BS distributions
for different combinations of $\{N,M\}$, and the main findings pertinent to our work 
are presented and discussed in the main sections of this paper. 
Another quantity that may be of interest in some cases is 
the maximum probability of a BS distribution for  a given unitary.  
This is determined mainly by the input state and the size 
of the Hilbert space.  As depicted in Fig. \ref{fig-A2},  the variations with the seed are rather small and thus 
to a good approximation we have 
\[
\max_{\bm r}\{P({\bm r}|{\bm s};\hat{\mathfrak U})\}\approx 1.52 |{\mathbb S}_{M,N}|^{-0.7}. 
\]
Moreover, it is worth emphasizing that according to our simulations, the maximum probability depends  
weakly on the actual values of $\{N,M\}$, but rather strongly on the total number of outcomes $|{\mathbb S}_{M,N}|$. More precisely,  
for two BS distributions over approximately the same number of outcomes, but for different 
combinations of $\{M,N\}$,  
the corresponding maximum probabilities are also approximately equal. 

\newpage

\newpage


\begin{thebibliography}{99}
\bibitem{remark1}  
Albeit rare, such events are unavoidable in practise and if necessary, they 
can be eliminated by means of post-selection i.e., 
their rejection, and the appropriate  renormalization of the final distribution over the collision-free configurations.
\bibitem{supercomputers}
For instance, IBM's Sequoia supercomputer system based at Lawrence Livermore National Laboratory 
delivers about $16\times 10^{15}$ flops, whereas Titan at Oak Ridge National Laboratory delivers about 
$18\times 10^{15}$ flops. The former has about 
$10^6$ CPU cores, while the latter is a hybrid system with 18688 16-core CPUs and 18688 Nvidia GPUs.
\end{thebibliography}
\end{document}